\documentstyle[epsfig]{mn2e}

\def\ApJ{{ApJ}}
\def\MNRAS{{ MNRAS}}
\def\aa{{ A\&A}}
\def\Nature{{ Nature}}
\def\GCN{{ GCN Circ}}
\def\PRD{{ Phys. Rev. D}}
\def\PRL{{ Phys. Rev. Lett}}
\def\etal{{\it et al}}

\def\be{\begin{equation}}
\def\ee{\end{equation}}
\def\bea{\begin{eqnarray}}
\def\eea{\end{eqnarray}}

\title[]{Gamma-ray Burst UV/optical afterglow polarimetry as a probe of Quantum Gravity}
\author[]{Yi-Zhong Fan$^{1,2}$\thanks{Lady Davis Fellow, E-mail: yzfan@pmo.ac.cn}, Da-Ming Wei$^{2,3}$
and Dong Xu$^4$\\
$^1${\sl The Racah Inst. of Physics, Hebrew University, Jerusalem 91904, Israel}\\
$^2$ {\sl Purple Mountain Observatory, Chinese Academy of
Science, Nanjing 210008, China}\\
$^3${\sl Joint Center for Particle Nuclear Physics and Cosmology
of Purple Mountain Observatory - Nanjing University, Nanjing
210008,China}\\
$^4${\sl Dark Cosmology Centre, Niels Bohr Institute, University
of Copenhagen, Juliane Maries Vej 30, 2100 Copenhagen, Denmark} }
\date{Accepted ......
Received ......; in original form ......}

\begin{document}

\maketitle
\begin{abstract}
A possible birefringence effect that arises in quantum gravity leads to a frequency-dependent
rotation of the polarization angle of linearly polarized emission from distant sources. Here
we use the UV/optical polarization data of the afterglows of GRB 020813 and GRB 021004 to
constrain this effect. We find an upper limit on the  Gambini \& Pulin birefringence parameter
$| \eta | <2\times 10^{-7}$. This limit is of 3 orders better than the previous limits from
observations of AGNs and of the Crab pulsar. Much stronger limits may be obtained by the
future observation of polarization of the prompt $\gamma$-rays.
\end{abstract}


Modifications to the standard dispersion relation of high energy
particles arise in various theories of quantum gravity
\cite{QG1,QG2,QG3}. These modifications result generically in a
modification to the standard dispersion relation. This, in turn,
introduces an energy dependent photon's velocity. Attempts to
explore this possibility and to set limit or observe it are
typically based on time of flight tests of very high energy
photons. In specific cases and in particular in loop quantum
gravity, the deviations from the standard dispersion relation take
a particular form in which photons with right handed and left
handed polarization travel with different velocities \cite{GP99}.
This leads  to a frequency dependent rotation of the polarization
vector of a linearly polarized light. Observations of a linear
polarization can, therefore, set limits on the magnitude of the
quantum gravity birefringence parameter
\cite{GP99,GK01,mitr03,Jac04,Kahn06}. The presence of linear
polarization in the UV/optical spectrum of distant galaxies and in
the the X-ray spectrum of the Crab Nebula sets upper bounds on the
birefringence parameter: $\eta \leq 10^{-4}$ \cite{GK01,Karr04}.
Much smaller limit, $\eta\leq 10^{-14}$, can be obtained using the
linear polarization detection in GRB 021206 \cite{mitr03,Jac04}.
However, the original claims of a detection of polarization of
this burst \cite{CB03} have been refuted convincingly
\cite{RF04,Wigg04}. The polarization analysis of GRB 030519B also
yielded null result \cite{Wigg04}. Possible evidence for linearly
polarized $\gamma-$rays in GRB 930131 and GRB 960924
\cite{Willis05} was also reported. If correct it will provide a
comparable upper limit. However, it is not clear yet how robust
are the novel methods used to estimate the polarization in these
bursts. We explore here the implications and limits that can be
set by observations of linear polarization from the optical
afterglow of GRB 020813 \cite{bart03} and GRB 021004
\cite{Lazz03}.

Gambini and Pullin (1999) have shown that within loop quantum gravity photons with left and
right circular polarizations will travel at different velocities $v_\pm =c(1\pm \eta \hbar
\omega/E_{_{\rm pl}})$, where $\pm$ denote the different polarization states and $\eta$ is a
dimensionless constant. Here we discuss the generalized form (see also Mitrofanov 2003 and the
references therein)
\begin{equation}
v_\pm =c[1\pm \eta (\hbar \omega/E_{_{\rm pl}})^n],
\label{eq:Mitr1}
\end{equation}
where $\eta^{-1/n}$ characterizes the scale of breakdown of the standard dispersion relation
in units of the Planck's energy $E_{_{\rm pl}}=1.22\times 10^{19}$ GeV. Naturally, one would
expect $\eta \sim 1$ corresponding to a typical energy scale of the Planck energy or slightly
higher, while a very low limit on $\eta$ would render such a modification unlikely.

Since the linear polarization is a superposition of two
monochromatic waves with opposite circular polarizations, the
plane of linear polarization is subject to a rotation along the
photons' path because of the difference between the two circular
components \cite{GP99,GK01,mitr03,Jac04}:
\begin{equation}
d\phi
=\eta ({\omega l_p \over c})^{n+1} {dL_{(z)}\over l_p},
\end{equation} where
$l_p=\sqrt{\hbar G/c^3}=\hbar c/E_{_{\rm pl}}\approx 1.6\times 10^{-33}$ cm is the Planck's
length scale and \[dL_{(z)}={c \over h H_0}{dz\over (1+z)\sqrt{\Omega_{\Lambda}+\Omega_{\rm M}(1+z)^{3}}}\] is the differential distance the
photons has travelled (for a flat cosmological model, i.e., $\Omega_{\rm M}+\Omega_\Lambda=1$), 
$h \sim 0.73$ is the current Hubble constant in units of
$H_0={100~\rm km/sec/Mpc}$.

\begin{figure}
\begin{center}
\includegraphics[width=100mm,height=80mm]{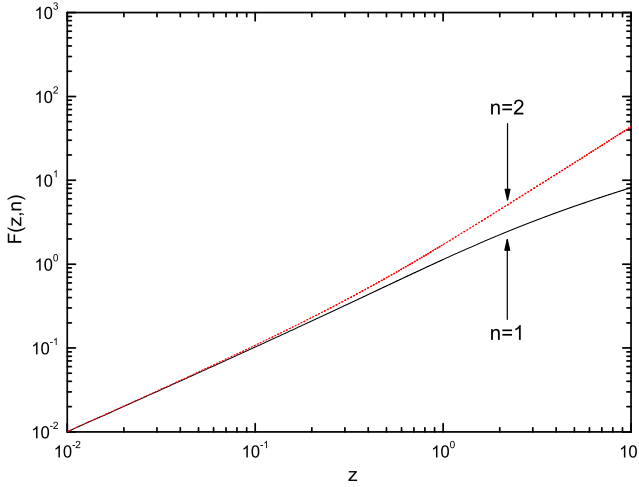}
\caption[...]{$F(z,n)$ in the redshift range $0.01-10$ for
$\Omega_{\rm M}=0.28$, $\Omega_\Lambda=0.72$, and $h=0.73$.}
\label{fig:Fz}
\end{center}
\end{figure}

The observed polarization angle at frequency $\nu_{\rm obs}$ of a
photon emitted at a redshift $z$ with an intrinsic polarization
angle $\Phi_0$ is
\begin{equation}
\Phi_{(n)} =\Phi_0+7.94\times 10^{60}~\eta {l_p^{n+1} \over
c^{n+1}} (2\pi \nu_{\rm obs})^{n+1} F(z,n).  \label{eq:Bas1}
\end{equation}
 The function $F(z,n)$ depends on the cosmology and can be expressed as 
\begin{equation}
 F(z,n) \equiv ({h\over 0.73})^{-1} \int^z_0 {(1+z)^n dz \over
\sqrt{\Omega_\Lambda+\Omega_{\rm M} (1+z)^3}}.
\end{equation}
The function $F \approx 1$ for $z=1$.
$F(z,1)\propto z$ for $z\ll 1$. $F(z,1)\propto z^{1/2}$ for $z\gg
1$, while $F(z,2)\propto z$ for $z\ll 1$ and $F(z,2)\propto
z^{3/2}$ for $z\gg 1$. The slow increase of $F(z,1)$ for $z>1$
 suggests that there is no advantage in going to very large
$z$ values.

For $n=1$, we have $\Delta \Phi_{(1)}=\Phi_{(1)}-\Phi_0=0.9 \eta
({\nu_{obs}\over 10^{12}{\rm Hz}})^2 F(z,1)$. Therefore, the
relevant frequencies for observing the birefringence effect are
$\nu_{obs} \geq \nu_{bir} \equiv 1.05\times 10^{12} \eta^{-1/2}
F(z,1)^{-1/2}$, where $\nu_{bir}$ is defined as $\Delta
\Phi_{(n)}(\nu_{bir})=1$. For $n=2$, $\Delta
\Phi_{(2)}=\Phi_{(2)}-\Phi_0=0.3
 \eta ({\nu_{obs}\over 10^{22}{\rm Hz}})^3
 F(z,2)$. So we have to probe the
birefringence effect with polarimetry of the $\sim 50$ MeV photons
from a source at $z\sim 1$. This is quite difficult because
usually at such high energy, there are not enough photons to allow
a reliable polarization measurement. Consequently, there was no
X$-$ray/$\gamma-$ray polarimeter planed for energy higher than
$30$ MeV \cite{mcco06}. In this work we  focus, therefore, on the
case of $n=1$.

At $\nu_{obs} \ll \nu_{bir}$ there is no significant observable
effect. On the other hand at $\nu_{obs} \gg \nu_{bir}$ the
rotation is so large that it will erase any intrinsic linear
polarization \cite{GK01,Jac04}. Therefore an detection of high
linear polarization at a frequency $\nu_{obs}$ from a source at
$z$ puts an upper limit on $\eta$:
\begin{equation}
\eta \leq ({\nu_{obs} \over 10^{12}~{\rm Hz}})^{-2}
F(z,1)^{-1},~{\rm or}~\eta \leq 3({\nu_{obs} \over 10^{22}~{\rm
Hz}})^{-3} F(z,2)^{-1}. \label{eq:Constr}
\end{equation}
In a limited frequency range around $\nu_{bir}$ we expect to
observe the birefringence effect as a frequency dependent linear
polarization vector. Such an observation could confirm the
existence of a birefringence effect and the corresponding quantum
gravity corrections provided that we can rule out an intrinsic
origin for the rotation of the polarization vector.

The UV/optical spectropolarimetry of GRB afterglows has been
reported in several events
\cite{wije99a,wije99b,bart03,wije99c,wije99d,wije99f,wije99e}.
Good quality results were obtained on GRB 020813 \cite{bart03} and
GRB 021004 \cite{Lazz03}. We use these two bursts to constrain the
possible birefringence effect.

GRB 020813 was detected by {\it High Energy Transient Explore 2}
(\emph{HETE-2}) on August 13rd 2002 \cite{Vill02} and found at the
redshift $z\sim 1.255$ \cite{bart03}. The UV/optical
[$(4000,~5000,~6300,~7300,~8300)\pm 500 $\AA] polarization
measurements of the afterglow of GRB 020813 were carried out at
$\sim 5.16,~6.27$, and 7.36 hours after the burst, respectively.
The observed polarization is in the range of $1.8\% -2.4\%$ (see
Table 2 of \cite{bart03} for details) and the scatter is very
small, in particular for the data collected at a given time. At
$t\sim 5.16$ hours, the corresponding polarization angles are
$(161\pm 1, 159 \pm1, 158\pm 1, 155\pm 1, 155\pm 1)$ deg,
respectively. At $t\sim 6.27$ hours, the corresponding
polarization angles are $(160\pm 1, 155 \pm1, 151\pm 1, 150\pm 1,
151\pm 2)$ deg, respectively. At $t\sim 7.36$ hours, the
corresponding polarization angles are $(153\pm 1, 149 \pm1, 156\pm
1, 153\pm 1, 149\pm 1)$ deg, respectively \cite{bart03}.

\begin{figure}
\begin{center}
\includegraphics[width=90mm,height=80mm]{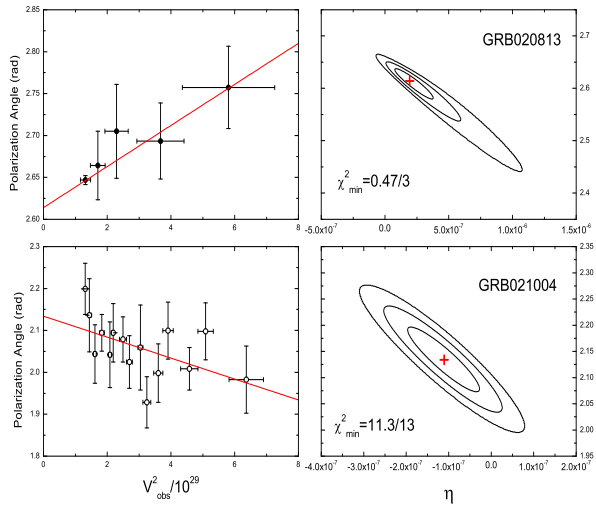}
\caption[...]{Fit to the afterglow polarimetry data of GRB 020813
(the upper panels; the angles are time-averaged) and GRB 021004
(the lower panels). The data are taken from \cite{bart03} and
\cite{Lazz03}, respectively. Left: Linear fit of the polarization
angles vs. the quantities of $\nu_{obs}^2/10^{29}$, taking double
errbars into account. Right: Joint confidence intervals
($1\sigma$, $2\sigma$, and $3\sigma$) in the $\Phi_0-\eta$ plane.
The cross indicates the best fit, corresponding to the minimal
chi-square value $\chi_{\rm min}^2$.} \label{fig:TimeAv}
\end{center}
\end{figure}

To constrain $\eta$ we look for a similar frequency dependent trend in this data. We allow a
variation in the polarization from one time to another and consider only the relative
polarization angle. To do so we subtract,  at each time, the average polarization angle (which
can vary intrinsically with time) and consider only the polarization angle with respect to
this average (see Fig.\ref{fig:TimeAv}).  The scatter in the shift between different times
provides an estimate of the error in the data. The best fitted $\eta =1.9 \times 10^{-7}$ and
$\eta=0$ is marginally consistent at the $2\sigma$ level (see
Fig.\ref{fig:TimeAv})\footnote{The direct time-averaged polarization data yields a less tight
constraint on the birefringence effect. In such a treatment, $\eta=0$ is consistent with the
data at $1\sigma$ level.} Upper limits at the $3 \sigma$ level that can be drawn from this
figure are: $ - 6 \times 10^{-8} < \eta < 1.2\times 10^{-6}$. These limits are two orders of
magnitude better than the limit obtained using optical polarization data of one distant galaxy
\cite{GK01} and X-ray polarimetry of the Crab Nebula \cite{Karr04}.

GRB 021004 was detected by \emph{HETE-2} \cite{Shir02} on October 4th 2002 and pinned down at
the redshift $z\sim 2.328$ by follow-up observations \cite{Mira02}. Here we directly take the
reduced spectropolarimetry data presented in Lazzati et al. (2003) using the Very Large
Telescope of the European Southern Observatory. As shown in the bottom left panel of Figure
\ref{fig:TimeAv}, we find the polarization angle of this burst is proportional to
$\nu_{obs}^2$ with a negative rather than a positive coefficient. The best fit indicates $\eta
=-1.1 \times 10^{-7}$. The dataset is consistent with the possibility of no shift at all
(i.e., $\eta=0$) at the $2\sigma$ confidence level.

\begin{figure}
\begin{center}
\includegraphics[width=90mm,height=80mm]{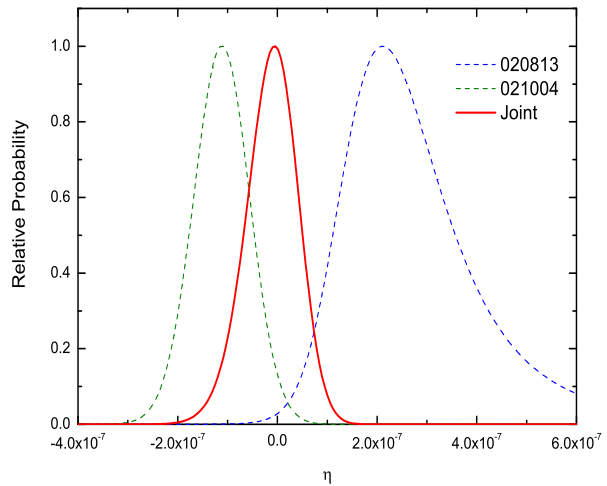}
\caption[...]{Individual and joint constraints on $\eta$ from the
UV/optical polarimetry of GRB 020813 and GRB 021004. For the joint
case, $-2\times 10^{-7}<\eta <1.4\times 10^{-7}$ at the $3\sigma$
confidence level.} \label{fig:Joint}
\end{center}
\end{figure}

Now we combine GRB020813 and GRB021004 to further investigate the possible birefringence
effect that arises in quantum gravity, that is, mainly to constrain the parameter $\eta$. To
realize this, we apply the Chi-Square fitting to the data of both bursts and marginalize the
Chi-Square merit function over all the possible polarization angles in order to derive the
probability only for the parameter $\eta$. The procedure is to (1) calculate the probability
for each given set $\eta$ and polarization angle for either burst; (2) integrate the
probabilities over polarization angle and then derive two average probabilities for each
$\eta$ for either burst; and (3) derive the joint probability for each $\eta$ for two bursts
by using the formula ${\rm{P(}}\eta _{\rm{i}} {\rm{)}} \propto {\rm{Exp}}[ - (\chi
_{020813,\eta _i }^2 + \chi _{020813,\eta _i }^2 )/2]$, where $i$ means the $i$-th $\eta$.
GRB020813 and GRB021004 are treated as independent events during the joint analysis. Shown in
Figure 3 are the likelihood functions for the parameter $\eta$ of GRB020318 (blue dash curve),
GRB021004 (olive dash curve), and their combination (red solid curve). At the 3$\sigma$
confidence level, we have $-2\times 10^{-7}\leq \eta \leq 1.4\times 10^{-7}$. This limit is
about 3 orders better than the previous estimates. For such stringent constraint on $\eta$,
the birefringence effect on our initial polarization measurement in a limited bandwidth ($500
$\AA ~for GRB 020813 and $200$\AA ~for GRB 021004) is unimportant and can be ignored.

It should be noted that the rotation of the polarization plane can also be induced by Faraday
rotation, so it is necessary to test whether the Faraday rotation could affect the results
discussed in this paper. It is well known that the rotation angle induced by the Faraday
rotation is $\Delta\alpha=8.1\times 10^{5}\lambda_m^{2}\int_{0}^{L} n_e B_{\|}dL$, where
$\lambda_m$ is the wavelength in units of meter, $n_e$ is the electron number density per
${\rm cm^{3}}$, $B_{\|}$ is the magnetic field strength (Gauss) parallel the propagating
direction, and $L$ is the distance in units of ${\rm pc}$. Typically, with $\lambda_m \sim
10^{-6}$, $n_e \sim 10^{-4}$, $B_{\|} \leq 10^{-12}$G and $L \sim 10^{11}$, we have
$\Delta\alpha \leq 10^{-11}$. As can be seen, even for $z\sim 10$, the rotation angle at
UV/optical and higher band is very small, therefore the Faraday rotation is insignificant.

A detection of polarization of the prompt emission of GRBs (at
such high frequency) will set immediately a much stronger limit on
$\eta$ \cite{GP99}. As mentioned before, a possible evidence for
the linearly polarized $\gamma-$rays ($\sim {\rm a ~few}$ hundred
keV) has been reported \cite{CB03} and ruled out
\cite{RF04,Wigg04} in GRB 021206. Wigger et al. \cite{Wigg04} also
suggest that there is no evidence for polarization of the prompt
$\gamma$-ray emission in GRB 030519B. On the other hand
polarization was detected in GRB 930131 and GRB 960924
\cite{Willis05} and possibly in GRB 041219a \cite{McGl06}. These
results are consistent with each other as (apart from the refuted
claims of \cite{CB03}) the errors are very large. The situation is
inconclusive and additional data is needed before the
corresponding strong limits on $\eta$ that have been suggested on
the basis of this burst \cite{Jac04} can be accepted. The planned
X-ray/$\gamma-$ray polarimitors POLAR \cite{prod05} or XPOL
\cite{cost06} could resolve this issue in the future. Even such
polarimeters won't help with a significant limit on $n=2$.  To
reach the limit $\eta =1$ for $n=2$ one needs a  polarization
measure of $10-100$ MeV signal from a cosmological distance.
However, even for the bursts with a hard,  highly-polarized
component as energetic as the one  detected in GRB 941017
\cite{Gonz03}, such a polarimetry requires a very large area
($\sim {\rm 1~m^2}$) which is beyond the goals of current design.

\section*{Acknowledgments}
We thank T. Piran and S. Covino for detailed comments, J. Hjorth for suggestions
and J. L. Han, D. Willis, S. McBreen, and X. F. Wu for
communications. DMW and YZF are supported by the National Natural
Science Foundation (grant 10673034) of China. DX is at the Dark
Cosmology Centre funded by The Danish National Research
Foundation.

\end{document}